\documentclass[]{article}

\usepackage[a4paper, left=0.6in, right=0.6in, top=0.6in, bottom=0.6in, footskip=.25in]{geometry}
\usepackage{authblk}
\usepackage{graphicx}
\usepackage{caption}

\usepackage{cite}

\title{Micro-scale Mechanical Metamaterial with a Controllable Transition in the Poisson's Ratio and Band Gap Formation}

\author[1,2]{K. K. Dudek*}
\author[1]{J. A. Iglesias Mart\'{\i}nez}
\author[1]{G. Ulliac}
\author[1]{L. Hirsinger}
\author[1]{L. Wang}
\author[1]{V. Laude}
\author[1]{M. Kadic*}

\affil[1]{Institut FEMTO-ST, CNRS, Université Bourgogne Franche-Comté, Besançon 25030, France} 
\affil[2]{Institute of Physics, University of Zielona Gora, ul. Szafrana 4a, Zielona Gora 65-069, Poland}

\date{}

\begin{document}
		
\maketitle



\begin{abstract}


The ability to change significantly mechanical and wave propagation properties of a structure without rebuilding it has been one of the main challenges in the field of mechanical metamaterials.
This stems from the enormous appeal that, especially in the case of micro-scale systems, such tunable behavior may offer from the perspective of applications ranging from biomedical to protective devices. In this work, a novel micro-scale mechanical metamaterial is proposed that can undergo a transition from one type of configuration to another, with one configuration having a very negative Poisson's ratio, corresponding to strong auxeticity, and the other having a highly positive Poisson's ratio. The formation of phononic band gaps, at the same time, can be controlled, which can be very useful in the design of vibration dampers and sensors. Finally, it is shown experimentally that reconfiguration of the system, leading to a change in its properties, can be induced and controlled remotely through application of a magnetic field, thanks to appropriately distributed magnetic inclusions.  

\end{abstract}


\section{Introduction}


Over the years, the importance of functional materials \cite{Florijn2014, Coulais_Teomy_2016, Coulais2018, Fleury2014_Science, Fleury2015_Nat_Commun, Wenz_2021, Zhu_Birla_2020, Huang_Jin_2020, Tao_2020, Jackson2018, Qi_review_2021}  has been rapidly increasing in our everyday lives. Most prominently in the span of the last two decades, different classes of these structures have been successfully implemented in numerous industries including biomedical \cite{Kolken_Zadpoor2018, Teunis_van_Manen_2021} and protective devices \cite{Miniaci_2016, Imbalzano_2016, Novak_Vesenjak_2022}, sports equipment \cite{Duncan2018} as well as electronics \cite{Russo2011, Huang_Liu_2019}. One of the most promising directions of studies related to such easily-applicable functional materials is that devoted to mechanical metamaterials \cite{Lakes1987, Evans_Alderson_2000, Gladman_2016, Overvelde_2017, Mirzaali_2018, Silverberg_Evans_2014, Rafsanjani2019, Cai_Wang_2021,  Song_Zou_2021, Li_Librandi_2021, Bertoldi_Vitelli_2017, Grima_triangles1, Neville_2016,chen2020optimal,chen2022closed,tan2022single, chen20223d, Deng_Bertoldi_2022}. Mechanical metamaterials are rationally designed structures that can exhibit counterintuitive mechanical properties based primarily on their design rather than their chemical composition. Some of the most commonly studied instances of such extraordinary mechanical properties are negative Poisson's ratio (auxetic behavior) \cite{Lakes1987, Wojciechowski_1989, Mizzi_Grima2015, Wei_Yang_2021, Wang_Adv_Sci_2022}, negative stiffness \cite{qu2017,Hewage2016, Dudek_Mater_Des_2020, Tan_Wang_2022} and negative compressibility \cite{qu2017,qu2018, Nicolau_Motter_2012, Baughman1998}. Most notably, auxetic mechanical metamaterials have been proven to exhibit superior indentation resistance \cite{Li_Liu_2020}, energy absorption \cite{Yuan_2018, Mohsenizadeh_2018} and wave attenuation \cite{Miniaci_2016, Krushynska_2017} in comparison to many conventional non-auxetic materials. However, despite the numerous advantages offered by mechanical metamaterials characterized by a negative value of Poisson's ratio, a vast majority of such structures share an important limitation. Namely, once typical auxetic metamaterials are manufactured, it is very difficult to change their mechanical characteristic without rebuilding the system. This, in turn, diminishes their applicability in the fields where variable response of the material is required. Nevertheless, as reported in recent years, this shortcoming can be addressed by the use of active mechanical metamaterials in a form of stimuli responsive structures.  \vspace{8pt}

Active mechanical metamaterials are structures that can exhibit a tunable mechanical behavior depending on the application of an external stimulus. 
In recent years, it has been demonstrated that in the case of multi-material structures, the mechanical properties of the system can be adjusted remotely in a process that does not require reconstructing the system. In the case of macroscopic systems, one of the most promising approaches corresponds to the use of appropriately distributed magnet-like inclusions that respond to the application of an external magnetic field \cite{Grima_Dudek_2013, Galea_Dudek_2021}. However, this task becomes much more challenging at small scales where one of the more feasible approaches seems to be the use of magnetic nanoparticles dispersed within a non-magnetic host material \cite{Jackson2018, Pishvar_2020, Montgomery_Wu_2021, Ma_Chang_2022}. Another approach making it possible to construct an active mechanical metamaterial is associated with the use of several materials with different thermal expansion coefficients. In such a case, the behavior of the structure can be modified by changing the external temperature \cite{Montgomery_Wu_2021, Ji_Johnny_2021}. This method, however, often does not allow to obtain a very significant variation in the mechanical properties. In addition, the corresponding reconfiguration process tends to be quite slow. 
Last but not least, it is worth noting that it is possible to construct an active metamaterial composed of only one type of residual material that, to some extent, can undergo a change in its mechanical properties with a deformation pattern based solely on a variation in the mechanical stimulus, e.g. the rate of the mechanical compression. A prime example of this approach corresponds to hierarchical metamaterials \cite{Gatt_hierarchical_2015, Cho_hierarchical_2014, Tang_Lin_hierarchical_2015, An_Bertoldi_2020, Billon_2016, Dudek_Adv_Mater_2022, Dudek_hierarch_pssb_2022}, where different hierarchical levels can normally deform irrespective of each other.
\vspace{8pt}

Despite several notable achievements, studies devoted to active mechanical metamaterials are still in their infancy and much more in-depth analysis must be conducted to fully harness their potential. First of all, it is important to emphasize that for a significant number of known active mechanical metamaterials, even though it is possible to change one of their properties without rebuilding the system, it is still very difficult to achieve control over the other properties. This, in turn, could increase the appeal for active functional materials from the perspective of many industries. In fact, even though this task is not simple, it becomes even more challenging at small scales, such as the micro-scale, where the manufacturing process is often not only very difficult but also fairly expensive. On the other hand, despite the aforementioned challenges, the ability to achieve tunable control over multiple properties of a micro-scale system without the need for reconstructing it appears to be a very appealing concept that could be utilized in fields such as programmable robotics, vibration dampers / sensors and effective biomedical devices. To this aim, a very promising idea seems to be the possibility of controlling the Poisson's ratio of a micro-scale system \cite{Dudek_Adv_Mater_2022} as well as its wave propagation properties with an emphasis on tunable band gap formation \cite{Krushynska_2017, Frenzel_Kopfler_2019, Mei_Li_2021}. In fact, in the literature \cite{Kunin_Yang_2016}, there are some rare examples demonstrating the possibility of controlling to a limited extent both Poisson's ratio and band gap formation of micro-scale mechanical metamaterials. However, despite their numerous advantages, such studies typically focus on complex hierarchical structures where the deformation process is not easily recoverable nor controlled due to the presence of multiple independent hierarchical levels. In many cases, it is also not possible to observe a large tunable change in the Poisson's ratio of the system, which would significantly increase the applicability of the considered concept.   \vspace{8pt}

In this work, a novel mechanical metamaterial is proposed that in general can be defined in the form of two distinct configurations.
It is presented that without rebuilding the system, the considered structure can undergo a transition from one type of configuration to another during which the Poisson's ratio varies very significantly.
Namely, the change occurs from a structure possessing a positive Poisson's ratio to a strongly auxetic structure and vice versa. In addition, in terms of wave propagation properties, it is shown that a change in the configuration of the system can be used to achieve control over the formation of a phononic band gap. Furthermore, it is presented that all configurations of the considered system can be constructed at the micro-scale where they exhibit the aforementioned tunable properties. Finally, it is shown experimentally that reconfiguration of the system, leading to a change in its properties, can be prompted and controlled remotely through the use of an external magnetic field.

\section{Results and Discussion}

\begin{figure}
	\centering
	\includegraphics[width=0.95\linewidth]{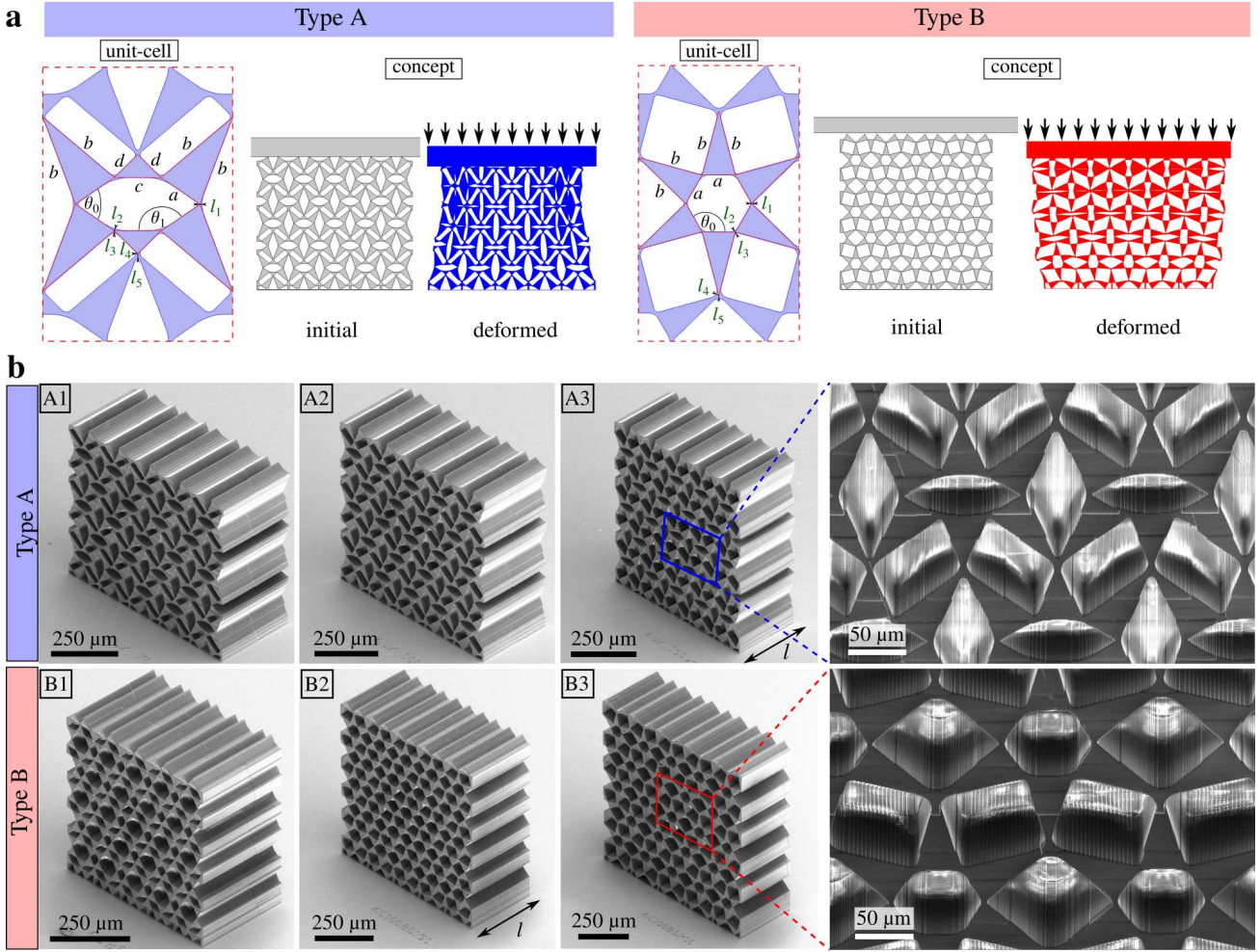}
	\caption{{\bf a)} Diagrams showing the two types of structures considered in this work. {\bf b)} Experimental prototypes corresponding to specific examples of the Type A and Type B structures that are analyzed from the point of view of their mechanical properties. Parameters that remain the same for all structures are the following: $a$ = 40 $\mu$m, $l_{1}$ = 4.8 $\mu$m, $l_{2}$ = 2.4 $\mu$m, $l_{3}$ = 2.4 $\mu$m, $l_{4}$ = 3.6  $\mu$m, and $l_{5}$ = 6.7 $\mu$m. In addition, parameters that remain the same for type A structures are $b$ = 80 $\mu$m and $c$ = 40 $\mu$m. Parameters that assume unique values both for type A and type B systems are as follows: $d$ = 28 $\mu$m (A1), $d$ = 40 $\mu$m (A2), $d$ = 60 $\mu$m (A3) and $b$ = 40 $\mu$m (B1), $b$ = 60 $\mu$m (B2), and $b$ = 80 $\mu$m (B3). The out-of-plane thickness $l$ for all of the structures is equal to 300 $\mu$m.  }
	\label{model}
\end{figure}

In this work, a novel mechanical metamaterial composed of triangle-shaped structural elements is proposed in order to assess its potential to exhibit versatile mechanical and wave propagation properties. As shown in {\bf Fig. \ref{model}a}, the considered system can occur in two very different configurations, that in this work, are named Type A and Type B structures. In the case of the Type A configuration, the system consists of two types of isosceles triangles connected to each other at vertices. Due to a specific connectivity, this structure corresponds to three different types of apertures. In contrast, the configuration defined as Type B has a much more regular design and consists of only one type of isosceles triangles. In fact, it resembles a lattice of mutually connected star-shaped elements where each of these elements consists of six triangles. The two considered configurations were designed in such a way that during mechanical deformation they would exhibit a very different Poisson's ratio. This concept is depicted schematically in {\bf Fig. \ref{model}a}, where qualitatively, type A and Type B structures exhibit strongly negative (auxetic behavior) and positive Poisson's ratio, respectively.  \vspace{8pt} 

In order to investigate the mechanical properties of the two considered configurations and to demonstrate that the proposed concept can be utilized in the case of microscopic applications, experimental prototypes were constructed at the micro-scale (see Methods section). In addition, as shown in Fig. {\bf \ref{model}b}, in order to assess the effect of the geometric parameters on the behavior of the structure, for each configuration three different prototypes were fabricated that are labeled A1, A2, A3 and B1, B2, B3. In fact, the only difference between the three Type A structures is the value of the $d$ parameter. Similarly, the only difference between the Type B prototypes is the value of the $b$ parameter. All prototypes are composed of 5 $\times$ 3 unit-cells, where the bottom part of the system is clamped as a result of the manufacturing process (see Methods section).   

\subsection{Mechanical Properties}

\begin{figure}
	\centering`
	\includegraphics[width=0.8\linewidth]{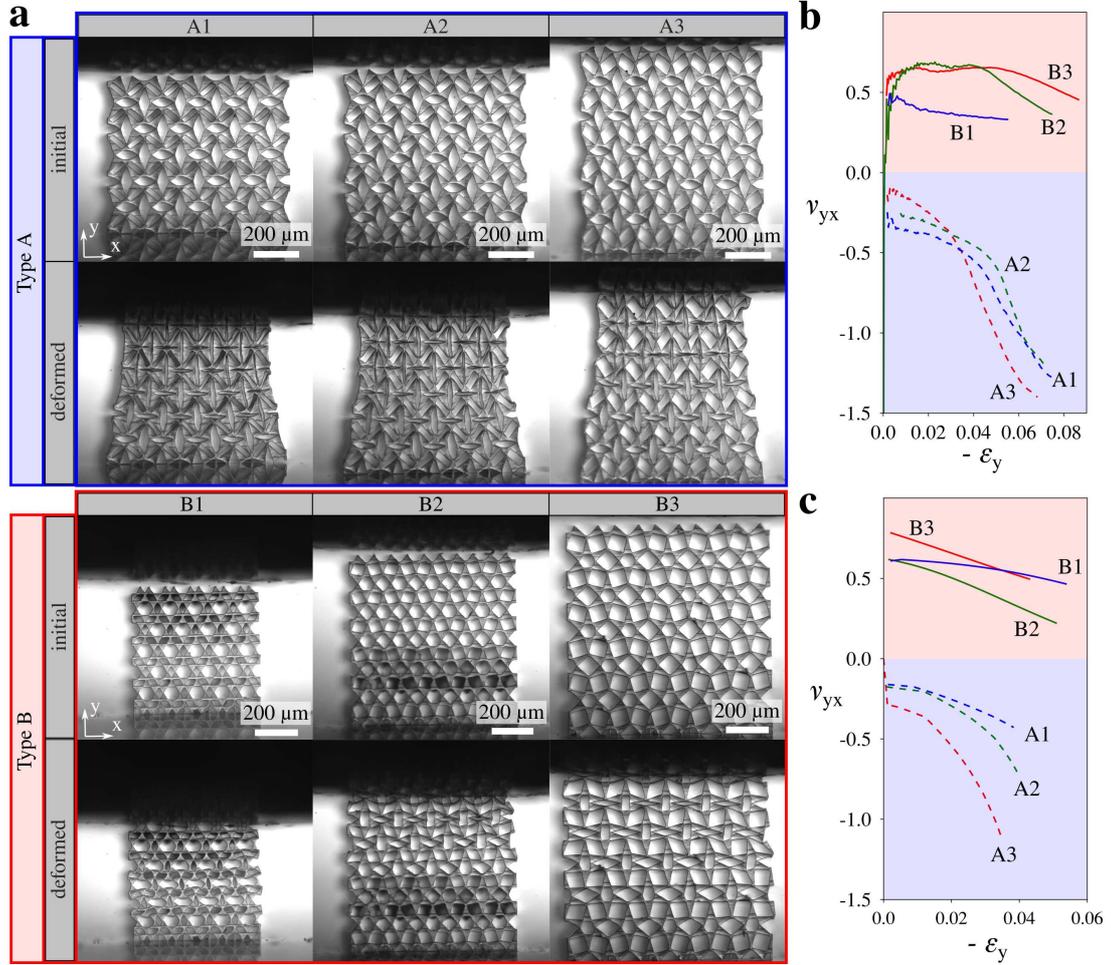}
	\caption{{\bf a)} Optical images of a compression test along the $y$-axis of the three micro-scale experimental prototypes corresponding to Type A and Type B configurations. {\bf b)} Poisson's ratio estimated from experimental samples. {\bf c)} Poisson's ratio estimated with FEM (Finite Element Method) computer simulations that was calculated based on the deformation of the central unit-cell of the considered finite samples.}
	\label{mechanical_results}
\end{figure}

To determine the Poisson's ratio of the considered system, the microscopic experimental prototypes corresponding both to Type A and Type B were compressed by a flat external indenter along the $y$-axis, with the bottom part of each sample remaining fixed in space. Thus, to qualitatively estimate the Poisson's ratio of the system, one can follow the changes in the horizontal dimension of the topmost part of the structure that is the farthest part of each sample from the aforementioned constraint. According to optical images of the experiment presented in {\bf Fig. \ref{mechanical_results}a}, one can note that from the qualitative analysis, prototypes corresponding to Type A and Type B exhibit a very different behavior. Namely, the $x$ dimension for all of the prototypes associated with the Type A configuration decreases significantly during the compression process. This, in turn, is an indication of a strong auxetic behavior. In contrast, in the case of structures corresponding to Type B, the horizontal dimension of the samples initially increases before returning approximately to the initial value at large strains. Such behavior is an indication of positive and close to zero Poisson's ratio, respectively. This means that Type B configurations exhibit a non-auxetic behavior unlike the Type A samples. At this point, it should be emphasized that non-uniform deformation of the entire system originates from the fact that the experimental samples remain attached to the substrate after the 3D printing process. However, if one was to consider the deformation of the detached sample, the extent of the non-uniformity would be significantly smaller. Nonetheless, mechanical testing could potentially become unreliable in such a scenario due to the very small size of the system. It is also worth emphasizing that if fabricated through the use of appropriate materials, the considered system can maintain its properties over numerous deformation cycles (see {\bf Fig. S1} in the {\bf Supplementary Information}). This, in turn, further enhances its appeal from the perspective of potential applications.   \vspace{8pt}

In addition to the qualitative description of the behavior of the system, it is important to conduct a quantitative analysis in order to determine the range of Poisson's ratio values and to assess its usefulness in the case of specific applications. To this aim, in order to calculate the Poisson's ratio for each of the structures, the central unit-cell was taken into consideration. This stems from the fact that the mechanical behavior of this specific unit-cell is the least affected by the factors contributing to the uneven deformation pattern. Namely, the deformation of this unit-cell is not significantly affected by the bottom part of the system being attached to the substrate. In addition, its behavior is also mostly unaffected by the edge effects on the left and right-hand sides of the structure as well as by the frition related to the contact of the topmost part of the system with the indenter inducing the deformation in the conducted experiments. As shown in {\bf Fig. \ref{mechanical_results}b}, all of the Type A structures exhibit a strong auxetic behavior throughout the entire deformation process. In fact, the largest extent of auxeticity can be observed for the A3 sample corresponding to the largest value of $d$. In this case, for a relatively large strain of around 0.07, the Poisson's ratio assumes the value of approximately -1.4. Furthermore, one can note that the extent of auxeticity can be adjusted by the value of $d$. In contrast, the behavior of the analyzed samples of type B is very different. Namely, all type B samples tend to exhibit a positive Poisson's ratio that in most cases, becomes gradually less positive as strain increases. Notably, the sample that exhibits the largest values of Poisson's ratio throughout the majority of the deformation process is structure B3 that possesses the largest value of $b$ amongst the considered samples. Thus, taking all of these results into account, there is a significant difference between the Poisson's ratios of type A and type B structures. In addition, the range of Poisson's ratio for each configuration can be adjusted by a variation in geometric parameters.   \vspace{8pt}

In addition to experimental studies, another interesting aspect of the analysis corresponds to the investigation of the Poisson's ratio of the system through the FEM (Finite Element Method) computer simulations. In this case, both the structure and the corresponding boundary conditions were designed to closely match the experiment (see Methods section). As shown in {\bf Fig. \ref{mechanical_results}c}, the trends observed for computational results were qualitatively the same as for conducted experiments. Namely, in the case of these two approaches, type A configurations exhibited a strong auxetic behavior throughout the entire deformation process while type B configurations exhibited positive values of this parameter. Of course, in terms of the quantitative analysis, there are some discrepancies between the two approaches that originate from multiple reasons. Namely, in the case of the experiment, the effect of the bottom constraint and edge effects on the effective Poisson's ratio is not fully negligible. However, through the use of multiple cells in the design of the system and the selection of only the central unit-cell for analysis purposes, these effects are certainly significantly diminished. Thus, based on the provided results, it is clear that the two approaches confirm a very significant difference in the Poisson's ratio exhibited by the two types of analyzed configurations.

\subsection{Effect of Reconfiguration on the Properties of the Structures}

In addition to the analysis of Poisson's ratio, another interesting direction of study is the analysis of phononic band gap formation associated with phonon dispersion. This stems from the fact that such analysis provides valuable information about the ranges of frequencies for which elastic waves are not transmitted through a given configuration of the system. Hence, in addition to applications related explicitly to a tunable Poisson's ratio, such information can be used in the design of microscopic vibration dampers / sensors. \vspace{8pt}

\begin{figure}
	\centering`
	\includegraphics[width=\linewidth]{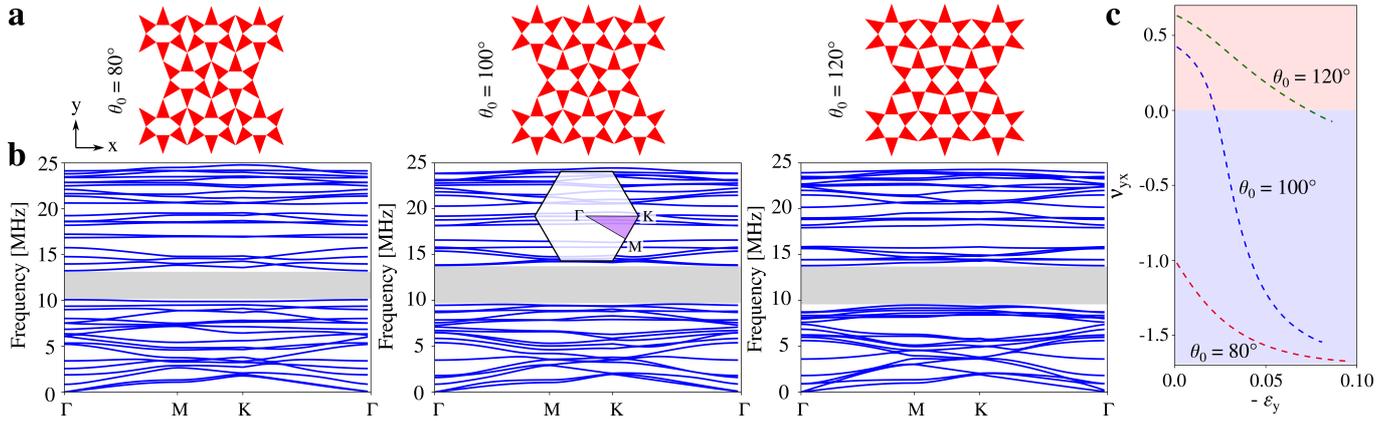}
	\caption{Effect of the reconfiguration of the system on phononic band gap formation and mechanical properties. {\bf a)} Diagrams showing schematically different configurations assumed by the same system. {\bf b)} Phonon dispersion relations for the three considered configurations where the first band gap is highlighted by means of the grey box. {\bf c)} Variation of Poisson's ratio for the three selected structures subjected to mechanical compression along the $y$-axis. Poisson's ratio was calculated for individual unit-cells with periodic boundary conditions imposed in the $x$-direction and a symmetric prescribed displacement applied to the topmost and bottommost parts of the system along the $y$-axis.}
	\label{reconfig_results}
\end{figure}

From the perspective of applications, it is very important to check what happens to mechanical and wave propagation properties once the considered system is subjected to a reconfiguration process. In other words, it is essential to assess the properties of the system during a reconfiguration process that does not involve reconstructing the structure and that could potentially occur as a result of the application of an external stimulus. An example of such a reconfiguration process is presented in {\bf Fig. \ref{reconfig_results}a}, where the initial configuration of type A corresponds to angle $\theta_{0} = 80^{\circ}$. In this case, all triangles are the same and correspond to the $b / a$ = 1.5 aspect ratio. Furthermore, as the reconfiguration process begins, the magnitude of the angle $\theta_{0}$ starts increasing until it reaches the value $\theta_{0} = 120^{\circ}$. At this point, the system assumes a configuration of type B, where there are only two types of apertures. To describe what happens to the properties of the system throughout such a reconfiguration process, we focus on the three representative configurations depicted schematically in {\bf Fig. \ref{reconfig_results}a}, where $\theta_{0}$ is equal to $80^{\circ}$, $100^{\circ}$ and $120^{\circ}$ respectively.      \vspace{8pt}

According to {\bf Fig. \ref{reconfig_results}b}, the phononic band structures for all three considered configurations are relatively similar. Most notably, the first non-negligible band gap occurring in the case of the system characterized by $\theta_{0} = 80^{\circ}$ has a size of 2.95 MHz. As the value of $\theta_{0}$ increases and assumes the values of $100^{\circ}$ and $120^{\circ}$, the size of the band gap increases and is equal to 4.03 MHz and 4.13 MHz respectively. However, even though the size of the band gap changes as a result of the reconfiguration process, one could say that such variations are not very significant. This, in turn, agrees with the observations made for other complex mechanical metamaterials subjected to a reconfiguration process that were reported in the literature \cite{Kunin_Yang_2016}. At this point, it is also important to note that whereas for Fig. \ref{reconfig_results} a structure corresponding to a specific value of the b/a ratio was taken into account, in general, the change in width of the first significant band gap may look different when considering the deformation of structures characterized by different values of $b/a$. More specifically, as described in the Supplementary Information, if one was to consider the deformation of the structure associated with $b/a$ = 2 from the configuration where $\theta_{0}$  = 80$^{\circ}$ to the configuration corresponding to $\theta_{0}$  = 120$^{\circ}$, the magnitude of the change in the band gap width would be very different from the former case. Namely, in this scenario, the relative change in the band gap width is only equal to 0.149 MHz. In contrast, for the structure associated with $b/a$ = 1.5, the change in the band gap width equals 1.18 MHz. Thus, a relatively small variation in the aspect ratio of structural elements may lead to a considerable difference in the magnitude of the change in the width of the band gap during the continuous reconfiguration process associated with the change in $\theta_{0}$. Of course, this stems from the fact that structures corresponding to certain values of $b/a$ lead to considerably different band gaps compared to other systems. Hence, it follows that the change in their magnitude is also considerably different. \vspace{8pt}

Contrary to the results related to band gap formation, it appears that the reconfiguration process strongly influences the Poisson's ratio of the system. As shown in {\bf Fig. \ref{reconfig_results}c}, configurations associated with $\theta_{0} = 80^{\circ}$ and $\theta_{0} = 120^{\circ}$ assume very different values of Poisson's ratio. Namely, in the case of the system where $\theta_{0} = 80^{\circ}$, the structure exhibits a strong auxetic behavior throughout the entire deformation process. On the other hand, for $\theta_{0} = 120^{\circ}$, the value of the Poisson's ratio is predominantly positive. These two results are in accordance with the expectations that one could have based on the results related to the mechanical testing of different configurations of Type A and Type B. However, it is very interesting to see what happens when the Type A configuration becomes very similar to a Type B configuration. In fact, such a scenario can be observed in the case of the intermediate configuration corresponding to $\theta_{0} = 100^{\circ}$. For this configuration, even though Poisson's ratio is initially positive, it quickly assumes strong negative values upon increasing strain. All of these results are very interesting since they indicate that a relatively simple reconfiguration process of the considered system makes it possible to very significantly change its Poisson's ratio.    \vspace{8pt}

\begin{figure}
	\centering
	\includegraphics[width=0.75\linewidth]{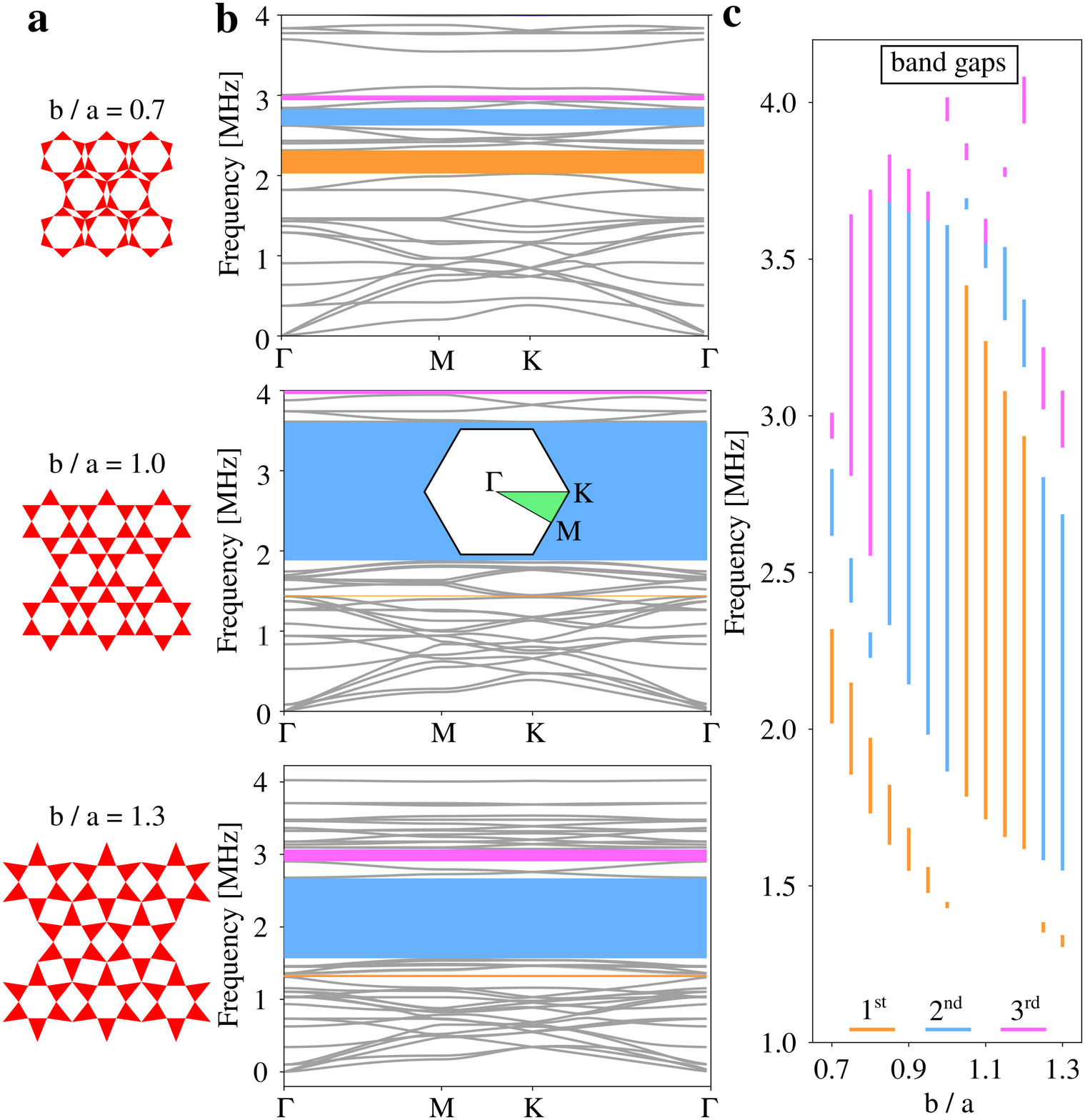}
	\caption{Effect of the elongation of the isosceles triangles corresponding to Type B of the system on the band gap formation. {\bf a)} Diagrams showing three different configurations of the system, where the value of $a$ is kept constant. {\bf b)} Phonon band structures corresponding to the analyzed structures. {\bf c)} The first three band gaps associated with structures corresponding to different values of the $b / a$ ratio. The geometric parameters used for all of the structures were set to be the following:  $a$ = 200 $\mu$m, $l_{1}$ = 8 $\mu$m, $l_{2}$ = 4 $\mu$m, $l_{3}$ = 4 $\mu$m, $l_{4}$ = 4  $\mu$m, $l_{5}$ = 9 $\mu$m}
	\label{scaling_of_triangles}
\end{figure}

Once the effect of the reconfiguration of the system on its properties has been determined, it is also interesting to have a closer look at band gap formation for the Type B system. As shown in {\bf Fig. \ref{reconfig_results}b}, the largest first significant band gap is observed during the reconfiguration process. To this aim, it was analyzed what happens when the aspect ratio of isosceles triangles constituting the Type B structure changes while keeping dimension $a$ constant. \vspace{8pt}

The three specific configurations of the system chosen for such analysis are portrayed schematically in {\bf Fig.\ref{scaling_of_triangles}a}. In fact, the only difference between them is the value of $b$. Hence, to refer to a specific configuration assumed by the structure, one can use a particular value of the $b / a$ coefficient. According to {\bf Fig.\ref{scaling_of_triangles}b}, the variation in the value of $b / a$ has a very significant effect on the phonon dispersion and on the size of the first significant band gap. More specifically, in the considered range of frequencies, one can note that for the structure associated with $b / a$ = 0.7, the first non-negligible band gap is relatively small. In contrast, once the geometric dimensions of the structure are modified to correspond to $b / a$ = 1.0, the appearance of a very significant band gap can be observed. Furthermore, upon increasing the magnitude of this coefficient up to the value of 1.3, the size of the first significant band gap is reduced relative to the configuration associated with $b / a$ = 1.0. This means that at relatively low frequencies, one should expect the largest band gap to be observed for this specific configuration. Hence, as shown in {\bf Fig. \ref{scaling_of_triangles}c}, an analysis of the size of the first three band gaps was conducted for a larger number of possible structures corresponding to the $b / a$ coefficient ranging between 0.7 and 1.3. Based on the results, the width of the first significant band gap gradually increases until it assumes its maximum value at $b / a$ = 1.0. For configurations with $b / a$ exceeding this value, the band gap gradually decreases. Finally, it should be noted that the experimental validation of the possibility of controlling the width of band gaps through the change in the value of $b/a$ is provided in {\bf Fig. S2} in the {\bf Supplementary Information}.

\subsection{Active Control}

\begin{figure}
	\centering
	\includegraphics[width=0.99\linewidth]{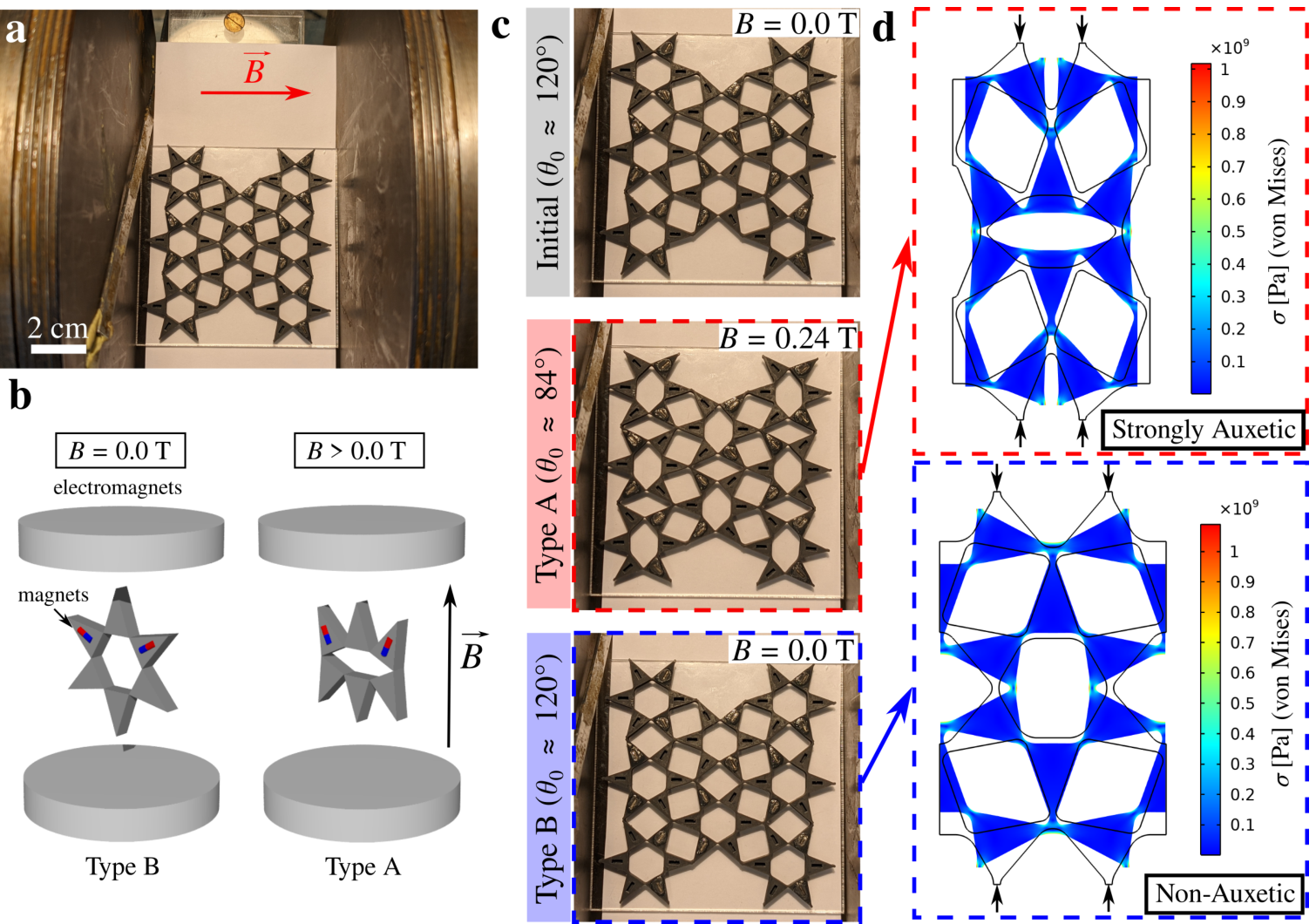}
	\caption{Active reconfiguration and transition in the properties of the system induced by the application of an external magnetic field. {\bf a)} Picture of the investigated sample placed between two parallel electromagnets that produce an approximately uniform magnetic field. {\bf b)} Diagram showing the concept of the reconfiguration of the magneto-mechanical structure. {\bf c)} A change in the configuration of the system induced by the variation in the magnitude of the external magnetic field. {\bf d)} Mechanical deformation of different configurations assumed by the system during the active reconfiguration.}
	\label{active_transition}
\end{figure}

Up to this point, it has been shown that the properties of the proposed mechanical metamaterial can be controlled as a result of the reconfiguration process. However, such reconfiguration was merely a theoretical concept and it was not explicitly demonstrated how it could be achieved without reconstructing the system. Thus, in this section, it is shown how such active control over the configuration assumed by the system and the resulting mechanical properties can be attained. Furthermore, since the concept proposed in this work is fully scalable and all results can in general be reproduced irrespective of the size of the system, active control over the reconfiguration process was demonstrated through the use of a macroscopic prototype (see {\bf Fig. \ref{active_transition}}).     \vspace{8pt}

As shown in {\bf Fig. \ref{active_transition}a}, the system selected to demonstrate the active reconfiguration process corresponds to a 3D-printed (see Methods section) prototype associated with the Type B configuration that was inserted into the uniform magnetic field induced by two large parallel electromagnets. The response of the structure to the application of the external magnetic field was made possible due to the use of magnetic inclusions in the form of very small neodymium magnets (see Methods section) that were embedded within some of the structural triangle-like elements constituting the system. The concept of active reconfiguration corresponding to this structure is schematically demonstrated in {\bf Fig. \ref{active_transition}b}, where for the sake of clarity, only a fragment of the considered structure is presented. As can be seen, the structure is initially at rest in the Type B configuration, where $\theta_{0}$ = 120$^{\circ}$. Once the external magnetic field is switched on, the magnetic dipole moments of the magnets attempt to align with the external magnetic field. Of course, depending on the magnitude of the external field, this might not be possible since for the triangles to rotate, the applied torques must be large enough to overcome the resistance and the restoring forces corresponding to the elastic non-magnetic material that the structure is made of.   \vspace{8pt}

According to {\bf Fig. \ref{active_transition}c}, upon being subjected to a relatively strong external magnetic field having a magnitude of 0.24 T, the considered structure deforms significantly and assumes a configuration corresponding to $\theta_{0} \approx 84^{\circ}$ (see {\bf Supplementary Video 1} to observe the gradual reconfiguration process where the magnetic field was changed with increments of 0.01 T). In other words, the structure undergoes a transition from Type B to Type A configuration. Based on the computational results provided in {\bf Fig. \ref{active_transition}d}, one can note that these two configurations correspond to a very different Poisson's ratio. Thus, the application of the external magnetic field makes it possible to significantly modify the properties of the structure without the need to rebuild it. In addition, after switching off the external magnetic field, the structure returns to its initial configuration, which means that after the transition it is possible to recover the initial properties of the system. These results are very important as they prove that the mechanical metamaterial proposed in this work can be used to exhibit a versatile Poisson's ratio that can be controlled remotely via the magnitude and orientation of the external magnetic field. Of course, in this case, due to the scale of the prototype, the controllable behavior was achieved thanks to the use of magnetic inclusions in the form of neodymium magnets. However, at much lower scales, this approach should be replaced for example by the use of single domain superparamagnetic nanoparticles dispersed within the resin  prior to the polymerization process \cite{Huang_Sakar_2016, Kim_Chung_2011}. Another approach that could be used to achieve control over the behavior of the system through the use of an external magnetic field, is to use a selective magnetic coating applied for example by sputtering following the 3D printing fabrication process \cite{Huang_Sakar_Mao_2015, Cui_Heyderman_2019}. At this point, we emphasize that the results of this work are scale independent. Depending on the size of the system, however, they can be better suited for specific applications.   \vspace{8pt}

The results show that the proposed system can exhibit two very useful properties at the same time that are further controllable remotely via an applied external magnetic field without the need for reconstructing the system. Specifically, it was shown that the considered structure can undergo a transition from positive to highly negative Poisson's ratio solely as a result of a reconfiguration process that can be invoked and controlled remotely and that does not involve rebuilding the system. In addition, after the transition, the system maintains a stable mechanical behaviour for relatively large strains. Such active and reversible control over mechanical properties, combined with the observation of a very significant variation of Poisson's ratio, is seldom observed in mechanical metamaterials and can be very useful for numerous applications. Furthermore, what makes these results even more appealing, is the fact that it is demonstrated that the highly versatile Poisson's ratio of the considered system can be observed for samples fabricated at the micro-scale. In addition, from the perspective of multifunctional devices, the proposed system exhibits two highly useful properties at once: controllable Poisson's ratio and adjustable phononic band gap. The latter property prevents waves in a specific range of frequencies from propagating through the system. Mechanical metamaterials exhibiting such band gap formation combined with fully controllable Poisson's ratio are extremely rare. In a few known instances where  control over both aspects was attempted to some extent \cite{Montgomery_Wu_2021}, the studies were limited to structures at the macro / millimetre scale and typically only one of these properties was quantitatively assessed. \vspace{8pt}

As mentioned above, the reconfiguration process leading to the transition in the properties of the system is fully reversible and can be achieved in an active manner that does not require rebuilding the system. This means that the ability of the mechanical metamaterial to exhibit a tunable Poisson's ratio could be used at the macroscale or the millimeter scale in the design of smart protective devices that could adjust their mechanical response depending on the specific impacting body, e.g. active car bumpers. Similarly, at the microscale, the controllable Poisson's ratio can lead to the design of materials characterized by programmable extent of their micro-indentation resistance. In addition, should somebody use magnetic inclusions to control the properties of the structure, then by distributing them in a non-uniform manner, it would be possible to observe a situation where the initially uniformly-designed structure exhibits a very different Poisson's ratio in different parts of the system. Such behavior, as demonstrated in another study \cite{Dudek_Adv_Mater_2022}, could enable the system to exhibit complex shape morphing upon being subjected to a mechanical deformation or through the application of an external magnetic field. Especially at the micro-scale, and among other applications, such shape morphing may be useful for the design of smart stents that would provide local support to specific parts of a blood vessel, where the location of the support could be changed throughout treatment. Finally, it was also demonstrated that in addition to control over the Poisson's ratio, one can form band gaps that are almost unaffected by the reconfiguration process. Thus, while changing the mechanical response of the sample, it would be possible to prevent waves corresponding to a specific range of frequencies from being transmitted through the system. Such ability could be very interesting from the perspective of the design of multifunctional vibration-damping protective devices. Alternatively, should the need arise, the band gap could be adjusted by a variation in geometric parameters of the system. In addition, for specific values of the $b/a$ parameter, the change in the width of the band gap induced by mechanical deformation, which process does not involve the reconstruction of the system, can be noticeable.

\section{Conclusion}

In this work, a novel micro-scale mechanical metamaterial was proposed. It was demonstrated that depending on the assumed configuration, the system can exhibit very different Poisson's ratios including strongly negative and positive values at relatively large strains. In addition, it was shown that such control over the mechanical behavior of the system can be achieved as a result of active reconfiguration, i.e. a process that does not require rebuilding the system. For such reconfiguration process, the transition in the properties of the structure is recoverable and can be controlled for example by means of an external magnetic field. Furthermore, in addition to control over Poisson's ratio, it was demonstrated that during the reconfiguration process one can approximately retain the band gap formation corresponding to the system. Alternatively, band gap formation can be also tailored by an appropriate variation in the geometric parameters. All of these results indicate that the proposed mechanical metamaterial can be very useful in a plethora of applications ranging from smart protective to biomedical devices.    


\section{Methods}

\subsection{Fabrication}

The Poisson's ratio of the considered system was assessed based on micro-scale experimental prototypes produced through the use of a commercial 3D printer (Photonic Professional GT+, Nanoscribe GmbH). The 3D printer used in this work operates based on the two-photon lithography method. All samples were printed with a resolution of around 2 $\mu$m which was possible thanks to use of the negative tone IP-S photoresin (Nanoscribe GmbH). Prior to the printing process, a drop of resin was deposited on an ITO-coated soda-lime glass substrate having the following dimensions: 25 × 25 × 0.7 mm$^{3}$. After deposition, specific fragments of the drop were photopolymerized by means of a 25X-objective with a femtosecond laser that was operating at $\lambda$ = 780 nm. Furthermore, some of the standard printing parameters were set to be the following: a galvanometric scanning speed of 100 mm/s, a laser power at the level of 90\% and slicing as well as hatching distances were set to be equal to 1 $\mu$m and 0.5 $\mu$m respectively. Once the printing process was finished, each sample was developed for 25 min in a solution of PGMEA (Propylene Glycol Methyl Ether acetate). This step of the procedure allows to remove the nonpolymerized  photoresist. Subsequently, the resulting structure was rinsed for a duration of three minutes in isopropyl alcohol (IPA) in order to remove the developer solution.   \vspace{8pt}

In the case of the macroscopic experimental prototype, the structure was 3D printed by means of a commercial FDM 3D-printer (Ultimaker) by means of the elastic TPU 95A material. Furthermore, in each of the star-shaped segments of the structure, there were two apertures with magnetic inclusions in the form of 3 neodymium magnets of the N42 type. Each of the magnets had a cylindrical shape with a height of 1 mm and a base diameter equal to 1 mm. The remanent induction of each magnet was estimated to be equal to 0.613 T. The geometric dimensions corresponding to the initial structure were set to be the following: $a$ = 6 mm, $b$ = 9 mm, $l_{1}$ = 0.7 mm, $l_{2}$ = 0.43 mm, $l_{3}$ = 0.36 mm, $l_{4}$ = 0.8 mm and $l_{5}$ = 0.55 mm.

\subsection{Mechanical Compression Testing}

The experimental samples considered in this work were tested mechanically in order to assess their Poisson's ratio. To this aim, all samples were subjected to the mechanical compression induced by a flat external indenter. The speed of the indenter was constant and was set to be equal to 1 $\mu$m s$^{-1}$. At the same time, the bottom part of each sample was fixed in space due to its connection to the substrate on top of which they were printed. Throughout the experiment, the deformation process was recorded by means of an optical camera with magnification factor x20. The working distance, i.e. the distance between the lens of the camera and the sample, was approximately equal to 11 mm. Based on the recorded pictures, the Poisson's ratio of the system was determined based on the central topmost unit-cell (see {\bf Supplementary Information}). To track the position of vertices that could be used to calculate the Poisson's ratio of the selected unit-cell, the digital image correlation method implemented in the MATLAB software was used.   \vspace{8pt}

In the case of the results presented in {\bf Fig. \ref{active_transition}}, the macroscopic experimental prototype was inserted into the uniform external magnetic field that was induced by two parallel electromagnets. The magnitude of the applied field was varied in a range between 0 T and 0.24 T with steps of 0.01 T in order to induce the reconfiguration process (see {\bf Supplementary Video 1}). The deformation of the structure was recorded by means of a standard optical camera.

\subsection{FEM simulations}

To determine the Poisson's ratio of the considered finite structurs resembling experimental prototypes presented in {\bf Fig. \ref{mechanical_results}}, FEM simulations were conducted by means of the commercial software COMSOL Multiphysics. To this aim, for each of the structures considered in {\bf Fig. \ref{mechanical_results}}, it was assumed that geometric parameters characterizing the structure are the same as in the case of the experimental samples. Furthermore, to match the constraint used in the experiments, it was assumed that the bottom part of each of the systems is fixed in space. In addition, the deformation of the structure was induced by means of contact with a large flat indenter in order to match the deformation process observed in the experiments as closely as possible. Mechanical properties of the simulated material were set as $E$ = 4 GPa, $\nu$ = 0.4 and $\rho$ = 1200 kg m$^{-3}$. In addition, geometrical nonlinearities were implemented to make simulations more realistic under large strains. Finally, it should be emphasized that a specific method related to the calculation of the Poisson's ratio is provided in the {\bf Supplementary Information}.  \vspace{8pt}

In order to determine the phonon band structure, Bloch boundary conditions were implemented and used via an eigenvalue solver assuming real wave vectors in the Bloch conditions and searching for real eigenfrequencies of the system.

\medskip
\noindent
\textbf{Supporting Information} \par 
Supporting Information is available from the Wiley Online Library or from the corresponding author upon request.

\medskip
\noindent
\textbf{Acknowledgements} \par 

This work was partly supported by the french RENATECH network and its FEMTO-ST technological facility.\\

K.K.D. acknowledges the support of the Polish National Science Centre (NCN) in the form of the grant awarded as a part of the SONATINA 5 program, project No. 2021/40/C/ST5/00007 under the name "Programmable magneto-mechanical metamaterials guided by the magnetic field".\\

This research was funded by the Polish Minister of Education and Science under the program “Regional Initiative of Excellence” in 2019-2023, project No. 003/RID/2018/19, funding amount PLN 11 936 596.10.

\medskip

\bibliographystyle{ieeetr}
\bibliography{manuscript_references.bib}

\begin{thebibliography}{10}

\bibitem{Florijn2014}
B.~Florijn, C.~Coulais, and M.~van Hecke, ``Programmable mechanical
  metamaterials,'' {\em Phys. Rev. Lett.}, vol.~113,, p.~175503, 2014.

\bibitem{Coulais_Teomy_2016}
C.~Coulais, E.~Teomy, K.~de~Reus, Y.~Shokef, and M.~van Hecke, ``Combinatorial
  design of textured mechanical metamaterials,'' {\em Nature}, vol.~535,,
  pp.~529--532, 2016.

\bibitem{Coulais2018}
C.~Coulais, A.~Sabbadini, F.~Vink, and M.~van Hecke, ``Multi-step self-guided
  pathways for shape-changing metamaterials,'' {\em Nature}, vol.~561,,
  pp.~512--515, 2018.

\bibitem{Fleury2014_Science}
R.~Fleury, D.~L. Sounas, C.~F. Fleck, M.~R. Haberman, and A.~Al\'{u}, ``Sound
  isolation and giant linear nonreciprocity in a compact acoustic circulator,''
  {\em Science}, vol.~343,, pp.~516--519, 2014.

\bibitem{Fleury2015_Nat_Commun}
R.~Fleury, D.~L. Sounas, and A.~Al\'{u}, ``An invisible acoustic sensor based
  on parity-time symmetry,'' {\em Nat. Commun.}, vol.~6,, p.~5905, 2015.

\bibitem{Wenz_2021}
F.~Wenz, I.~Schmidt, A.~Leichner, T.~Lichti, S.~Baumann, H.~Andrae, and
  C.~Eberl, ``Designing shape morphing behavior through local programming of
  mechanical metamaterials,'' {\em Adv. Mater.}, vol.~33,, p.~2008617, 2021.

\bibitem{Zhu_Birla_2020}
Y.~Zhu, M.~Birla, K.~R. Oldham, and E.~T. Filipov, ``Elastically and
  plastically foldable electrothermal micro-origami for controllable and rapid
  shape morphing,'' {\em Adv. Funct. Mater.}, vol.~30,, p.~2003741, 2020.

\bibitem{Huang_Jin_2020}
T.-Y. Huang, H.-W. Huang, D.~D. Jin, Q.~Y. Chen, J.~Y. Huang, L.~Zhang, and
  H.~L. Duan, ``Four-dimensional micro-building blocks,'' {\em Sci. Adv.},
  vol.~6,, 2020.

\bibitem{Tao_2020}
H.~Tao and J.~Gibert, ``Multifunctional mechanical metamaterials with embedded
  triboelectric nanogenerators,'' {\em Adv. Funct. Mater.}, vol.~30,,
  p.~2001720, 2020.

\bibitem{Jackson2018}
J.~A. Jackson, M.~C. Messner, N.~A. Dudukovic, W.~L. Smith, L.~Bekker,
  B.~Moran, A.~M. Golobic, A.~J. Pascall, E.~B. Duoss, K.~J. Loh, and C.~M.
  Spadaccini, ``Field responsive mechanical metamaterials,'' {\em Sci. Adv.},
  vol.~4,, p.~eaau6419, 2018.

\bibitem{Qi_review_2021}
J.~Qi, Z.~Chen, P.~Jiang, W.~Hu, Y.~Wang, Z.~Zhao, X.~Cao, S.~Zhang, R.~Tao,
  Y.~Li, and D.~Fang, ``Recent progress in active mechanical metamaterials and
  construction principles,'' {\em Adv. Sci.}, p.~2102662, 2021.

\bibitem{Kolken_Zadpoor2018}
H.~M.~A. Kolken, S.~Janbaz, S.~M.~A. Leeflang, K.~Lietaert, H.~H. Weinans, and
  A.~A. Zadpoor, ``Rationally designed meta-implants: A combination of auxetic
  and conventional meta-biomaterials.,'' {\em Mater. Horiz.}, vol.~5,,
  pp.~28--35, 2018.

\bibitem{Teunis_van_Manen_2021}
T.~van Manen, S.~Janbaz, K.~M.~B. Jansen, and A.~A. Zadpoor, ``4d printing of
  reconfigurable metamaterials anddevices,'' {\em Commun. Mater.}, vol.~2,,
  p.~56, 2021.

\bibitem{Miniaci_2016}
M.~Miniaci, A.~Krushynska, F.~Bosia, and N.~M. Pugno, ``Large scale mechanical
  metamaterials as seismic shields,'' {\em New J. Phys.}, vol.~8,, p.~083041,
  2016.

\bibitem{Imbalzano_2016}
G.~Imbalzano, P.~Tran, T.~D. Ngo, and P.~V.~S. Lee, ``A numerical study of
  auxetic composite panels under blast loadings,'' {\em Compos. Struct.},
  vol.~135,, pp.~339--352, 2016.

\bibitem{Novak_Vesenjak_2022}
N.~Novak, M.~Borovinsek, O.~Al-Ketan, Z.~Ren, and M.~Vesenjak, ``Impact and
  blast resistance of uniform and graded sandwich panels with tpms cellular
  structures,'' {\em Compos. Struct.}, vol.~300,, p.~116174, 2022.

\bibitem{Duncan2018}
O.~Duncan, T.~Shepherd, C.~Moroney, L.~Foster, P.~D. Venkatraman, K.~Winwood,
  T.~Allen, and A.~Alderson, ``Review of auxetic materials for sports
  applications: Expanding options in comfort and protection,'' {\em Appl.
  Sci.}, vol.~8,, p.~941, 2018.

\bibitem{Russo2011}
A.~Russo, B.~Y. Ahn, J.~J. Adams, E.~B. Duoss, J.~T. Bernhard, and J.~A. Lewis,
  ``Pen-on-paper flexible electronics,'' {\em Adv. Mater.}, vol.~23,,
  pp.~3426--3430, 2011.

\bibitem{Huang_Liu_2019}
S.~Huang, Y.~Liu, Y.~Zhao, Z.~Ren, and C.~F. Guo, ``Flexible electronics:
  Stretchable electrodes and their future,'' {\em Adv. Funct. Mater.},
  vol.~29,, p.~1805924, 2019.

\bibitem{Lakes1987}
R.~Lakes, ``Foam structures with a negative poisson's ratio,'' {\em Science},
  vol.~235,, p.~1038, 1987.

\bibitem{Evans_Alderson_2000}
K.~E. Evans and A.~Alderson, ``Auxetic materials: Functional materials and
  structures from lateral thinking!,'' {\em Adv. Mater.}, vol.~12,,
  pp.~617--628, 2000.

\bibitem{Gladman_2016}
A.~S. Gladman, E.~A. Matsumoto, R.~G. Nuzzo, L.~Mahadevan, and J.~A. Lewis,
  ``Biomimetic 4d printing,'' {\em Nat. Mater.}, vol.~15,, pp.~413--418, 2016.

\bibitem{Overvelde_2017}
J.~T.~B. Overvelde, J.~C. Weaver, C.~Hoberman, and K.~Bertoldi, ``Rational
  design of reconfigurable prismatic architected materials,'' {\em Nature},
  vol.~541,, pp.~347--352, 2017.

\bibitem{Mirzaali_2018}
M.~J. Mirzaali, S.~Janbaz, M.~Strano, L.~Vergani, and A.~A. Zadpoor,
  ``Shape-matching soft mechanical metamaterials,'' {\em Sci. Rep.}, vol.~8,,
  p.~965, 2018.

\bibitem{Silverberg_Evans_2014}
J.~L. Silverberg, A.~A. Evans, L.~McLeod, R.~C. Hayward, T.~Hull, C.~D.
  Santangelo, and I.~Cohen, ``Using origami design principles to fold
  reprogrammable mechanical metamaterials,'' {\em Science}, vol.~345,,
  pp.~647--650, 2014.

\bibitem{Rafsanjani2019}
A.~Rafsanjani, K.~Bertoldi, and A.~R. Studart, ``Programming soft robots with
  flexible mechanical metamaterials,'' {\em Sci. Robot.}, vol.~4,, p.~eaav7874,
  2019.

\bibitem{Cai_Wang_2021}
P.~Cai, C.~Wang, H.~Gao, and X.~Chen, ``Mechanomaterials: A rational deployment
  of forces and geometries in programming functional materials,'' {\em Adv.
  Mater.}, vol.~33,, p.~2007977, 2021.

\bibitem{Song_Zou_2021}
C.~Song, B.~Zou, Z.~Cui, Z.~Liang, and J.~Ju, ``Thermomechanically triggered
  reversible multi-transformability of a single material system by energy
  swapping and shape memory effects,'' {\em Adv. Funct. Mater.}, vol.~31,,
  p.~2101395, 2021.

\bibitem{Li_Librandi_2021}
S.~Li, G.~Librandi, Y.~Yao, A.~J. Richard, A.~Schneider-Yamamura, J.~Aizenberg,
  and K.~Bertoldi, ``Controlling liquid crystal orientations for programmable
  anisotropic transformations in cellular microstructures,'' {\em Adv. Mater.},
  vol.~33,, p.~2105024, 2021.

\bibitem{Bertoldi_Vitelli_2017}
K.~Bertoldi, V.~Vitelli, J.~Christensen, and M.~van Hecke, ``Flexible
  mechanical metamaterials,'' {\em Nat. Rev. Mater.}, vol.~2,, p.~17066, 2017.

\bibitem{Grima_triangles1}
J.~N. Grima, R.~Gatt, B.~Ellul, and E.~Chetcuti, ``Auxetic behaviour in
  non-crystalline materials having star or triangular shaped perforations,''
  {\em J. Non. Cryst. Solids}, vol.~356,, pp.~1980--1987, 2010.

\bibitem{Neville_2016}
R.~M. Neville, F.~Scarpa, and A.~Pirrera, ``Shape morphing kirigami mechanical
  metamaterials,'' {\em Sci. Rep.}, vol.~6,, p.~31067, 2016.

\bibitem{chen2020optimal}
X.~Chen, J.~Moughames, Q.~Ji, J.~A.~I. Mart{\'\i}nez, H.~Tan, S.~Adrar,
  N.~Laforge, J.-M. Cote, S.~Euphrasie, G.~Ulliac, {\em et~al.}, ``Optimal
  isotropic, reusable truss lattice material with near-zero poisson’s
  ratio,'' {\em Extreme Mechanics Letters}, vol.~41, p.~101048, 2020.

\bibitem{chen2022closed}
X.~Chen, Q.~Ji, J.~A.~I. Martinez, H.~Tan, G.~Ulliac, V.~Laude, and M.~Kadic,
  ``Closed tubular mechanical metamaterial as lightweight load-bearing
  structure and energy absorber,'' {\em Journal of the Mechanics and Physics of
  Solids}, vol.~167, p.~104957, 2022.

\bibitem{tan2022single}
X.~Tan, J.~A.~I. Mart{\'\i}nez, G.~Ulliac, B.~Wang, L.~Wu, J.~Moughames,
  M.~Raschetti, V.~Laude, and M.~Kadic, ``Single-step-lithography micro-stepper
  based on frictional contact and chiral metamaterial,'' {\em Small}, vol.~18,
  no.~28, p.~2202128, 2022.

\bibitem{chen20223d}
X.~Chen, J.~Moughames, Q.~Ji, J.~A.~I. Mart{\'\i}nez, H.~Tan, G.~Ulliac,
  V.~Laude, and M.~Kadic, ``3d lightweight mechanical metamaterial with nearly
  isotropic inelastic large deformation response,'' {\em Journal of the
  Mechanics and Physics of Solids}, vol.~169, p.~105057, 2022.

\bibitem{Deng_Bertoldi_2022}
B.~Deng, A.~Zareei, X.~Ding, J.~C. Weaver, C.~H. Rycroft, and K.~Bertoldi,
  ``Inverse design of mechanical metamaterials with target nonlinear response
  via a neural accelerated evolution strategy,'' {\em Adv. Mater.}, vol.~34,,
  p.~2206238, 2022.

\bibitem{Wojciechowski_1989}
K.~W. Wojciechowski, ``Two-dimensional isotropic system with a negative poisson
  ratio,'' {\em Phys. Lett. A}, vol.~137,, pp.~60--64, 1989.

\bibitem{Mizzi_Grima2015}
L.~Mizzi, K.~M. Azzopardi, D.~Attard, J.~N. Grima, and R.~Gatt, ``Auxetic
  metamaterials exhibiting giant negative poisson's ratios,'' {\em Phys. Status
  Solidi RRL}, vol.~9,, pp.~425--430, 2015.

\bibitem{Wei_Yang_2021}
Y.-L. Wei, Q.-S. Yang, and R.~Tao, ``Smp-based chiral auxetic mechanical
  metamaterial with tunable bandgap function,'' {\em Int. J. Mech. Sci.},
  vol.~195,, p.~106267, 2021.

\bibitem{Wang_Adv_Sci_2022}
L.~Wang, G.~Ulliac, B.~Wang, J.~A.~I. Martinez, K.~K. Dudek, V.~Laude, and
  M.~Kadic, ``3d auxetic metamaterials with elastically-stable continuous phase
  transition,'' {\em Adv. Sci.}, p.~2204721, 2022.

\bibitem{qu2017}
J.~Qu, M.~Kadic, and M.~Wegener, ``Poroelastic metamaterials with negative
  effective static compressibility,'' {\em Applied Physics Letters}, vol.~110,
  no.~17, p.~171901, 2017.

\bibitem{Hewage2016}
T.~A.~M. Hewage, K.~L. Alderson, A.~Alderson, and F.~Scarpa, ``Double-negative
  mechanical metamaterials displaying simultaneous negative stiffness and
  negative poisson’s ratio properties,'' {\em Adv. Mater.}, vol.~28,,
  pp.~10323--10332, 2016.

\bibitem{Dudek_Mater_Des_2020}
K.~K. Dudek, R.~Gatt, and J.~N. Grima, ``3d composite metamaterial with
  magnetic inclusions exhibiting negative stiffness and auxetic behaviour,''
  {\em Mater. Des.}, vol.~187,, p.~108403, 2020.

\bibitem{Tan_Wang_2022}
X.~Tan, L.~Wang, S.~Zhu, S.~Chen, B.~Wang, and M.~Kadic, ``A general strategy
  for performance enhancement of negative stiffness mechanical metamaterials,''
  {\em Eur. J. Mech. A Solids}, vol.~96,, p.~104702, 2022.

\bibitem{qu2018}
J.~Qu, M.~Kadic, and M.~Wegener, ``Three-dimensional poroelastic metamaterials
  with extremely negative or positive effective static volume
  compressibility,'' {\em Extreme Mechanics Letters}, vol.~22, pp.~165--171,
  2018.

\bibitem{Nicolau_Motter_2012}
Z.~G. Nicolaou and A.~E. Motter, ``Mechanical metamaterials with negative
  compressibility transitions,'' {\em Nat. Mater.}, vol.~11,, pp.~608--613,
  2012.

\bibitem{Baughman1998}
R.~H. Baughman, S.~Stafstrom, C.~Cui, and S.~O. Dantas, ``Materials with
  negative compressibilities in one or more dimensions,'' {\em Science},
  vol.~279,, pp.~1522--1524, 1998.

\bibitem{Li_Liu_2020}
T.~Li, F.~Liu, and L.~Wang, ``Enhancing indentation and impact resistance in
  auxetic composite materials,'' {\em Compos. B. Eng.}, vol.~198,, p.~108229,
  2020.

\bibitem{Yuan_2018}
S.~Yuan, C.~K. Chua, and K.~Zhou, ``3d-printed mechanical metamaterials with
  high energy absorption,'' {\em Adv. Mater. Technol.}, vol.~4,, p.~1800419,
  2019.

\bibitem{Mohsenizadeh_2018}
M.~Mohsenizadeh, F.~Gasbarri, M.~Munther, A.~Beheshti, and K.~Davami,
  ``Additively-manufactured lightweight metamaterials for energy absorption,''
  {\em Mater. Des.}, vol.~139,, pp.~521--530, 2018.

\bibitem{Krushynska_2017}
A.~O. Krushynska, M.~Miniaci, F.~Bosia, and N.~M. Pugno, ``Coupling local
  resonance with bragg band gaps in single-phase mechanical metamaterials,''
  {\em Extreme Mech. Lett.}, vol.~12,, pp.~30--36, 2017.

\bibitem{Grima_Dudek_2013}
J.~N. Grima, R.~Caruana-Gauci, M.~R. Dudek, K.~W. Wojciechowski, and R.~Gatt,
  ``Smart metamaterials with tunable auxetic and other properties,'' {\em Smart
  Mater. Struct.}, vol.~22,, p.~084016, 2013.

\bibitem{Galea_Dudek_2021}
R.~Galea, K.~K. Dudek, P.-S. Farrugia, L.~Z. Mangion, J.~N. Grima, and R.~Gatt,
  ``Reconfigurable magneto-mechanical metamaterials guided by magnetic
  fields,'' {\em Compos. Struct.}, vol.~280,, p.~114921, 2022.

\bibitem{Pishvar_2020}
M.~Pishvar and R.~L. Harne, ``Foundations for soft, smart matter by active
  mechanical metamaterials,'' {\em Adv. Sci.}, vol.~7,, p.~2001384, 2020.

\bibitem{Montgomery_Wu_2021}
S.~M. Montgomery, S.~Wu, X.~Kuang, C.~D. Armstrong, C.~Zemelka, Q.~Ze,
  R.~Zhang, R.~Zhao, and H.~H.~J. Qi, ``Magneto-mechanical metamaterials with
  widely tunable mechanical properties and acoustic bandgaps,'' {\em Adv.
  Funct. Mater.}, vol.~31,, p.~2005319, 2021.

\bibitem{Ma_Chang_2022}
C.~Ma, Y.~Chang, S.~Wu, and R.~R. Zhao, ``Deep learning-accelerated designs of
  tunable magneto-mechanical metamaterials,'' {\em ACS Appl. Mater.
  Interfaces}, vol.~14,, pp.~33892--33902, 2022.

\bibitem{Ji_Johnny_2021}
Q.~Ji, J.~Moughames, X.~Chen, G.~Fang, J.~J. Huaroto, V.~Laude, J.~A.~I.
  Martinez, G.~Ulliac, C.~Clevy, P.~Lutz, K.~Rabenorosoa, V.~Guelpa,
  A.~Spangenberg, J.~Liang, A.~Mosset, and M.~Kadic, ``4d thermomechanical
  metamaterials for soft microrobotics,'' {\em Commun. Mater.}, vol.~2,, p.~93,
  2021.

\bibitem{Gatt_hierarchical_2015}
R.~Gatt, L.~Mizzi, J.~I. Azzopardi, K.~M. Azzopardi, D.~Attard, A.~Casha,
  J.~Briffa, and J.~N. Grima, ``Hierarchical auxetic mechanical
  metamaterials,'' {\em Sci. Rep.}, vol.~5,, p.~8395, 2015.

\bibitem{Cho_hierarchical_2014}
Y.~Cho, J.-H. Shin, A.~Costa, V.~K. T.~A.~Kim, J.~Li, S.~Y. Lee, S.~Yang, H.~N.
  Han, I.-S. Choi, and D.~J. Srolovitz, ``Engineering the shape and structure
  of materials by fractal cut,'' {\em Proc. Natl. Acad. Sci.}, vol.~111,,
  pp.~17390--17395, 2014.

\bibitem{Tang_Lin_hierarchical_2015}
Y.~Tang, G.~Lin, L.~Han, S.~Qiu, S.~Yang, and J.~Yin, ``Design of
  hierarchically cut hinges for highly stretchable and reconfigurable
  metamaterials with enhanced strength,'' {\em Adv. Mater.}, vol.~27,,
  pp.~7181--7190, 2015.

\bibitem{An_Bertoldi_2020}
N.~An, A.~G. Domel, J.~Zhou, A.~Rafsanjani, and K.~Bertoldi, ``Programmable
  hierarchical kirigami,'' {\em Adv. Funct. Mater.}, vol.~30,, p.~1906711,
  2020.

\bibitem{Billon_2016}
K.~Billon, I.~Zampetakis, F.~Scarpa, M.~Ouisse, E.~Sadoulet-Reboul, M.~Collet,
  A.~Perriman, and A.~Hetherington, ``Mechanics and band gaps in hierarchical
  auxetic rectangular perforated composite metamaterials,'' {\em Compos.
  Struct.}, vol.~160,, pp.~1042--1050, 2017.

\bibitem{Dudek_Adv_Mater_2022}
K.~K. Dudek, J.~A.~I. Mart\'{\i}nez, G.~Ulliac, and M.~Kadic, ``Micro-scale
  auxetic hierarchical mechanical metamaterials for shape morphing,'' {\em Adv.
  Mater.}, vol.~34,, p.~2110115, 2022.

\bibitem{Dudek_hierarch_pssb_2022}
K.~K. Dudek, J.~A.~I. Mart\'{\i}nez, and M.~Kadic, ``Variable dual auxeticity
  of the hierarchical mechanical metamaterial composed of re-entrant structural
  motifs,'' {\em Phys. Status Solidi B}, p.~2200404, 2022.

\bibitem{Frenzel_Kopfler_2019}
T.~Frenzel, J.~K{\"o}pfler, E.~Jung, M.~Kadic, and M.~Wegener, ``Ultrasound
  experiments on acoustical activity in chiral mechanical metamaterials,'' {\em
  Nat. Commun.}, vol.~10,, p.~3384, 2019.

\bibitem{Mei_Li_2021}
C.~Mei, L.~Li, H.~Tang, X.~Han, X.~Wang, and Y.~Hu, ``Broadening band gaps of
  shear horizontal waves of metamaterials via graded hierarchical
  architectures,'' {\em Compos. Struct.}, vol.~271,, p.~114118, 2021.

\bibitem{Kunin_Yang_2016}
V.~Kunin, S.~Yang, Y.~Cho, P.~Deymier, and D.~J. Srolovitz, ``Static and
  dynamic elastic properties of fractal-cut materials,'' {\em Extreme Mech.
  Lett.}, vol.~6,, pp.~103--114, 2016.

\bibitem{Huang_Sakar_2016}
H.-W. Huang, M.~S. Sakar, A.~J. Petruska, S.~Pane, and B.~J. Nelson, ``Soft
  micromachines with programmable motility and morphology,'' {\em Nat.
  Commun.}, vol.~7, p.~12263, 2016.

\bibitem{Kim_Chung_2011}
J.~Kim, S.~E. Chung, S.-E. Choi, H.~Lee, J.~Kim, and S.~Kwon, ``Programming
  magnetic anisotropy in polymeric microactuators,'' {\em Nature Mater.},
  vol.~10, pp.~747--752, 2011.

\bibitem{Huang_Sakar_Mao_2015}
T.-Y. Huang, M.~S. Sakar, A.~Mao, A.~J. Petruska, F.~Qiu, X.-B. Chen,
  S.~Kennedy, D.~Mooney, and B.~J. Nelson, ``3d printed microtransporters:
  Compound micromachines for spatiotemporally controlled delivery of
  therapeutic agents,'' {\em Adv. Mater.}, vol.~27, pp.~6644--6650, 2015.

\bibitem{Cui_Heyderman_2019}
J.~Cui, T.-Y. Huang, Z.~Luo, P.~Testa, H.~Gu, X.-Z. Chen, B.~J. Nelson, and
  L.~J. Heyderman, ``Nanomagnetic encoding of shape-morphing micromachines,''
  {\em Nature}, vol.~575, pp.~164--168, 2019.

\end{thebibliography}

\end{document}